\documentclass[aps,prl,showpacs,twocolumn,amsmath,amssymb]{revtex4}

\usepackage{amsmath}
\usepackage{amsfonts}
\usepackage{graphicx}
\usepackage{dcolumn}
\usepackage{bm}

\begin{document}

\title{Mechanisms for Lasing with Cold Atoms as the Gain Medium}

\author{William Guerin}
\author{Franck Michaud}
\author{Robin Kaiser}
 \email{Robin.Kaiser@inln.cnrs.fr}
\affiliation{Institut Non Lin\'eaire de Nice, CNRS and Universit\'e de Nice Sophia-Antipolis,\\
  1361 route des Lucioles, 06560 Valbonne, France.}

\date{\today}

\begin{abstract}
{We realize a laser with a cloud of cold rubidium atoms as gain
medium, placed in a low-finesse cavity. Three different regimes of
laser emission are observed corresponding respectively to Mollow,
Raman and Four Wave Mixing mechanisms. We measure an output power
of up to $300~\mu$W and present the main properties of these
different lasers in each regime.}
\end{abstract}




\maketitle

Since Letokhov's seminal paper \cite{Letokhov:1968}, random lasers
have received increasing interest. Random lasing occurs when the
optical feedback due to multiple scattering in the gain medium
itself is sufficiently strong to reach the lasing threshold. In
the past decade, it has been observed in a variety of systems (see
\cite{Wiersma:2008} for a review) but many open questions remain
to be investigated, for which better characterized samples would
be highly valuable. A cloud of cold atoms could provide a
promising alternative medium to study random lasing, allowing for
a detailed understanding of the microscopic phenomena and a
precise control of essential parameters such as particle density
and scattering cross section. These properties have been exploited
to study coherent backscattering of light \cite{Labeyrie:1999} and
radiation trapping \cite{RadiationTrapping} in large clouds of
cold atoms. As many different gain mechanisms have been observed
with cold atoms, combining multiple scattering and gain in cold
atomic clouds seems a promising path towards the realization of a
new random laser. Besides the realization of a random laser, cold
atoms might allow to study additional features, such as the
transition from superfluorescence \cite{Rydberg} to amplified
spontaneous emission \cite{Boyd:1987} in a multiple scattering
regime. One preliminary step along this research lines is to use a
standard cavity to trigger laser oscillation with cold atoms as
gain medium. Such a laser may also be an interesting tool for
quantum optics, as one can take advantage of the nonlinear
response of the atoms to explore nonclassical correlations or
obtain squeezing \cite{Slusher:1985}.

In this letter, we present the realization of a cold-atom laser,
that can rely on three different gain mechanisms, depending on the
pumping scheme. By pumping near resonance, Mollow gain
\cite{Mollow:1972,Wu:1977} is the dominant process and gives rise
to a laser oscillation, whose spectrum is large (of the order of
the atomic natural linewidth), whereas by pumping further from
resonance, Raman gain between Zeeman sublevels \cite{Grison:1991}
gives rise to a weaker, spectrally sharper laser
\cite{Hilico:1992}. At last, by using two counter-propagating pump
beams, degenerate four-wave mixing (FWM)
\cite{Yariv:1977,Abrams:1978} produces a laser with a power up to
$300~\mu$W. By adjusting the atom-laser detuning
or the pump geometry, we can
continuously tune the laser from one regime to another.

Our experiment uses a cloud of cold $^{85}$Rb atoms confined in a
vapor-loaded Magneto-Optical Trap (MOT) produced by six large
independent trapping beams, allowing the trapping of up to
$10^{10}$ atoms at a density of $10^{10}$~atoms/cm$^3$,
corresponding to an on-resonance optical thickness of about 10. A
linear cavity, formed by two mirrors (a coupling-mirror with
curvature $RC1=1$~m, reflection coefficient $R1=0.95$ and plane
end mirror with reflection coefficient $R2\approx0.995$) separated
by a distance $L=0.8$~m is placed outside the vacuum chamber,
yielding a large round  trip loss $\mathcal{L}=32\%$ with a
correspondingly low finesse $\mathcal{F}=16$. The waist of the
fundamental mode of the cavity at the MOT location is
$w_\mathrm{cav}\approx\,500~\mu$m. To add gain to our system, we
use either one or two counter-propagating pump beams, denoted F
(forward) and B (backward), produced from the same laser with a
waist $w_\mathrm{pump}=2.6$~mm, with linear parallel polarizations
and a total available power of $P=80$~mW, corresponding to a
maximum pump intensity of $I=2P/(\pi
w_\mathrm{pump}^2)\approx\,750$~mW/cm$^2$. The pump is tuned near
the $F=3\rightarrow F'=4$ cycling transition of the $D2$ line of
$^{85}$Rb (frequency $\omega_\mathrm{A}$, wavelength $\lambda =
780$~nm, natural linewidth $\Gamma/2\pi = 5.9$~MHz), with an
adjustable detuning $\Delta=\omega_\mathrm{F,B}-\omega_\mathrm{A}$
and has an incident angle of $\approx 20^\circ$ with the cavity
axis. An additional beam P is used as a local oscillator to
monitor the spectrum of the laser or as a weak probe to measure
single-pass gain (insets of Figs. 2-4) with a propagation axis
making an angle with the cavity axis smaller than $10^\circ$. Its
frequency $\omega_\mathrm{P}$ can be swept around the pump
frequency with a detuning
$\delta=\omega_\mathrm{P}-\omega_\mathrm{F,B}$. Both lasers, pump
and probe, are obtained by injection-locking of a common master
laser, which allows to resolve narrow spectral features. In our
experiments, we load a MOT for 29~ms, and then switch off the
trapping beams and magnetic field gradient during 1ms, when lasing
or pump-probe spectroscopy are performed. In order to avoid
optical pumping into the dark hyperfine $F=2$ ground state, a
repumping laser is kept on all the time. Data acquisitions are the
result of an average of typically 1000 cycles.

As in a conventional laser, lasing occurs if gain exceeds losses
in the cavity, which can be observed as strong directional light
emission from the cavity. As we will discuss in detail below, we
are able to produce lasing with cold atoms as gain medium using
three different gain mechanisms: Mollow gain, Raman gain and Four
Wave Mixing (FWM). We can control the different mechanisms by the
pump geometry and the pump detuning $\Delta$ (see Table
\ref{table}). Mollow and Raman gain mechanisms only require a
single pump beam (F), whereas FWM only occurs when both pump beams
F and B are present and carefully aligned. With a single pump
beam, we find Mollow gain to be dominating close to the atomic
resonance, whereas Raman gain is more important for detunings
larger than $|\Delta|\approx\,4\Gamma$. Furthermore, the different
gain mechanisms lead to distinct polarizations. Mollow gain
generates a lasing mode with a polarization parallel to the pump
polarization, because the Mollow amplification is maximum for a
field aligned with the driven atomic dipole \cite{Mollow:1972}. On
the contrary, different polarizations between the pumping and the
amplified waves are necessary to induce a Raman transition between
two Zeeman substates: the polarization of the Raman laser is thus
orthogonal to the pump polarization. Lastly, the FWM laser has a
more complex polarization behaviour, as it is orthogonal for red-,
and parallel for blue-detuned pumps. We have checked that for any
pump detuning or probe power, the weak-probe FWM reflectivity is
stronger for orthogonal probe polarization, as expected from
previous experiments and models \cite{Lezama:2000}. We speculate
that pump-induced mechanical effects \cite{Gattobigio:2006} or
more complex collective coupling between the atoms and the cavity
\cite{Ritsch:2006} might be the origin of this polarization
behavior.

\begin{table}[t]
\caption{Different regimes of cold-atom laser versus pump
detuning. The polarization of the lasers are either parallel
($\parallel$) or orthogonal ($\perp$) to the polarization of the
pump beams.}\label{table}
\begin{tabular}{cccc} \hline \hline
  Pump beam(s) & \hspace{2pt} $\Delta<-4\Gamma$ &  \hspace{2pt}$-4\Gamma<\Delta<+4\Gamma$  &  \hspace{2pt}$\Delta>+4\Gamma$ \\
       \hline
F &   \hspace{5pt}Raman ($\perp$)  &  \hspace{5pt}Mollow ($\parallel$) &  \hspace{5pt}Raman ($\perp$)  \\
F+B &   \hspace{5pt}FWM ($\perp$)  &  \hspace{5pt}Mollow ($\parallel$) &  \hspace{5pt}FWM ($\parallel$)  \\
\hline \hline
\end{tabular}
\medskip
\end{table}

\begin{figure}[t]
\centering
\includegraphics{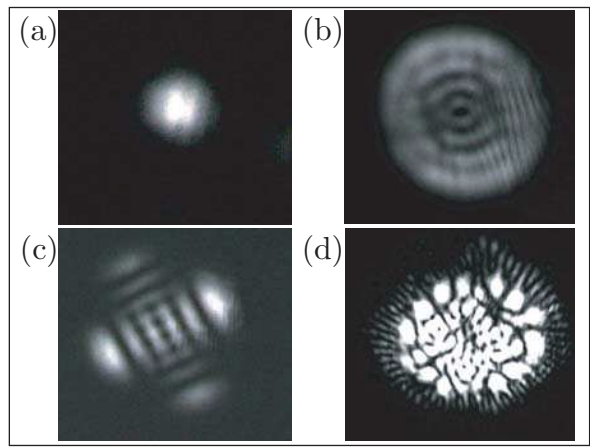}
\caption{Transverse modes of cold-atom lasers. (a) Gaussian
TEM$_{00}$ mode, obtained by inserting a small diaphragm in the
cavity. Typical modes of (b) the Mollow laser, (c) the Raman
laser, and (d) the four-wave mixing laser.} \label{fig.laser}
\end{figure}

In Fig. \ref{fig.laser} we show spatial (transverse) patterns of
these lasers, observed by imaging the beam onto a CCD camera.
Without any spatial filtering in the cavity, the different lasers
(Mollow, Raman and FWM) yield distinct transverse patterns. In
Fig. \ref{fig.laser}(b) [Fig. 1(c)] we show the transverse pattern
obtained with a Mollow (Raman) laser. We note that the Mollow
laser typically produces transverse patterns with radial
symmetries well described by Laguerre-Gauss modes, whereas the
modes of the Raman laser are rather Hermite-Gauss modes. The
origin of such radial or Cartesian symmetry may arise from the
different polarization of those two lasers: the radial symmetry is
preserved for the Mollow laser polarization and is broken for the
Raman laser one, probably due to slightly different losses in the
cavity. Fig. \ref{fig.laser}(d) shows the transverse pattern of
the FWM laser. As phase conjugation mechanisms are at work in such
a laser, any transverse mode can easily cross the lasing threshold
and complex lasing patterns are produced \cite{Lind:1981}.

We now turn to a more detailed description of the gain mechanisms
of the different lasers. The quantitative understanding of their
behavior needs to take into account effects such as pump geometry
and parameters (intensity, detuning), gain spectra, gain
saturation and mechanical effects induced by the pump beam(s).

\begin{figure}[!bt]
\centering
\includegraphics{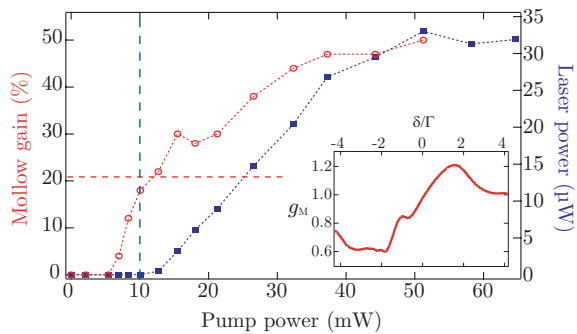} \caption{Laser power (squares) and Mollow gain
(open circles) versus pump power, with $b_0=11$ and $\Delta= +
\Gamma$. Lasing threshold (vertical dashed line) is expected to
appear with a gain of about $21\%$ (horizontal dashed line), in
good agreement with the experimental data. Inset: Typical
weak-probe transmission spectrum.} \label{fig.mollow}
\end{figure}

Let us first discuss the Mollow laser. Amplification of a weak
probe beam can happen when a two-level atom is excited by one
strong pump beam \cite{Mollow:1972, Wu:1977}. The corresponding
single-pass gain is $g_\mathrm{M}= \exp [-b_0
f_\mathrm{M}(\Omega,\Delta,\delta)]$, where $b_0$ is the
on-resonance optical thickness (without pump) of the cold-atom
cloud. The expression of $f_\mathrm{M}(\Omega,\Delta,\delta)$ can
be obtained from Optical Bloch Equations \cite{Mollow:1972}:
\begin{equation}\label{Eq.mollowfunction}
\begin{split}
f_\mathrm{M}&(\Omega,\Delta,\delta) =
\frac{\Gamma}{2}\,\frac{|z|^2}{|z|^2+\Omega^2/2} \times
\\
& \text{Re}
\left[\frac{(\Gamma+i\delta)(z+i\delta)-i\Omega^2\delta/(2z)}{(\Gamma+i\delta)(z+i\delta)(z^\ast+i\delta)+\Omega^2(\Gamma/2+i\delta)}
\right] \, ,
\end{split}
\end{equation}
where $z = \Gamma/2 - i\Delta$ and $\Omega$ is the Rabi frequency
of the atom-pump coupling, related to the pump intensity $I$ by
$\Omega^2 = \mathcal{C}^2 \Gamma^2 I/(2I_\mathrm{sat})$
($I_\mathrm{sat}=1.6$~mW/cm$^2$ is the saturation intensity and
$\mathcal{C}$ is the averaged Clebsch-Gordan coefficient of the
$F=3\rightarrow F'=4$ transition for a linear polarization). In
our setup we observe single-pass gain higher than $50\%$, with a
large gain curve (width $>\Gamma$). The shape of the transmission
spectrum (inset of Fig. \ref{fig.mollow}) is consistent with Eq.
(\ref{Eq.mollowfunction}). From Eq. (\ref{Eq.mollowfunction}) we
can also predict the maximum gain in respect to the pump
parameters $\Omega$, $\Delta$. We observe good agreement between
the behavior of the laser power and of the function $f_\mathrm{M}$
when varying $\Delta$: the maximum gain and laser power are
achieved for $|\Delta| \sim 2 \Gamma$ (the exact value depends on
$\Omega$) and $\Delta =0$ is a local minimum. However, we measured
a lower maximum gain than predicted by Eq.
(\ref{Eq.mollowfunction}). This is due to gain-saturation induced
by re-scattering of spontaneous emission inside the atomic cloud
\cite{Khaykovich:1999}.

As shown in Fig. \ref{fig.mollow} (squares), we observe a Mollow
laser emission with an output intensity reaching 35 $\mu$W. Taking
into account the round-trip losses $\mathcal{L}$, the condition
for laser oscillation is $g_\mathrm{M}^2 (1-\mathcal{L})
>1$. This corresponds to a gain at threshold of
$g_\mathrm{M}=1.21$ (horizontal line in Fig. \ref{fig.mollow}), in
good agreement with the observation.

\begin{figure}[!bt]
\centering
\includegraphics{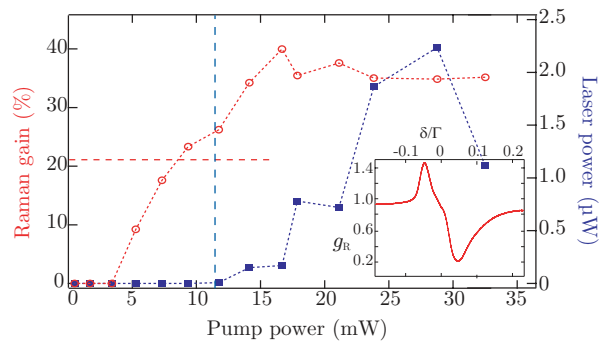}
\caption{Laser power (squares) and Raman gain (open circles)
versus pump power, with $b_0=10$ and $\Delta= -7 \Gamma$. Lasing
threshold (vertical dashed line) is expected to appear with a gain
of about 21 $\%$ (horizontal dashed line), in good agreement with
the experimental data. Inset: Typical weak-probe transmission
spectrum.} \label{fig.raman}
\end{figure}

When the pump frequency in detuned farther away from the atomic
resonance, Raman gain becomes dominant. Raman gain relies on the
pump-induced population inversion among the different
light-shifted $m_F$ Zeeman sublevels of the $F=3$ hyperfine level
\cite{Grison:1991,Brzozowski:2005}. Single-pass Raman gain of a
weak probe can be written $g_\mathrm{R}= e^{-b_0
f_\mathrm{R}(\Omega,\Delta,\delta)}$. For $|\Delta|\gg\Gamma$,
$f_\mathrm{R}(\Omega,\Delta,\delta)$ is given by
\begin{equation}
f_\mathrm{R}
=-\frac{\Omega^2}{\Delta^{2}}\left(\frac{A_1}{(\delta+\delta_R)^{2}+\gamma^{2}/4}-\frac{A_2}{(\delta-\delta_R)^{2}+\gamma^{2}/4}\right)\;
,
\end{equation}
where $A_{1,2}$ are the respective weights of the amplification
and absorption, $\delta_\mathrm{R}$ is the frequency difference
between the Zeeman sublevels and $\gamma$ is the width of the
Raman resonance \cite{Brzozowski:2005}. We have observed the laser
spectrum with a beat-note experiment, and we have checked that its
frequency corresponds to the maximum gain and is related to the
differential pump-induced light-shift $\delta_\mathrm{R}$ of the
different Zeeman sublevels. The width of the Raman resonance
$\gamma$ is related to the elastic scattering rate of the pump
photons and is much lower than $\Gamma$, due to the strong
detuning $\Delta$. The result is thus a much narrower gain
spectrum than in the previous case (inset of Fig.
\ref{fig.raman}). This leads to an important practical limitation
of the single-pumped Raman laser: atoms are pushed by the pump
beam, acquiring a velocity $v$, and the subsequent Doppler shift
becomes quickly larger than the width of the gain spectrum. As a
consequence, the gain in the cold-atom cloud is no longer the same
for a wave copropagating with the pump beam (F) and the wave
running in the counterpropagating direction. For the copropagating
direction, the relative Doppler shift is negligible, whereas for
the counterpropagating wave, a Doppler shift of $\sim
2\omega_\mathrm{A} v/c$, larger than the width of the gain
spectrum, leads to a suppression of the corresponding gain. As a
consequence, emission of our Raman laser stops after
$\approx\,20~\mu$s \cite{footnote}.

In Fig. \ref{fig.raman} we plot the output power of the Raman
laser as a function of pump power. A comparison with the
single-pass gain $g_{R}$ is again in good agreement for the
threshold condition  $g_\mathrm{R}^2 (1-\mathcal{L}) >1$ : for
Raman gain above $21\%$ laser emission occurs. As shown in Fig.
\ref{fig.raman} (squares), the output power of the Raman laser
emission ($\approx2~\mu$W) is much lower than the Mollow laser
one. This lower output power might arise from a lower saturation
intensity for Raman gain \cite{Michaud:2007}. Nevertheless, with a
weak signal, the Raman gain can be as high as $g_\mathrm{R}=2$
\cite{Michaud:2007}.

\begin{figure}[!bt]
\centering
\includegraphics{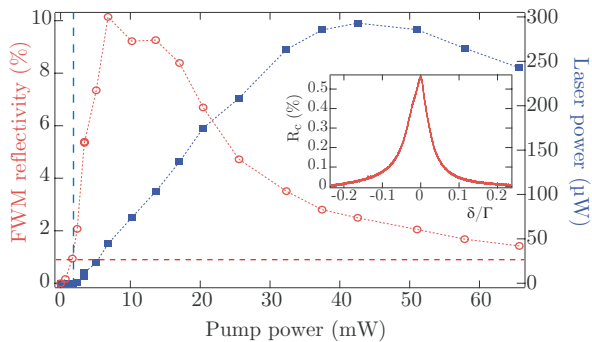}
\caption{Laser power (squares) and phase-conjugate reflectivity
due to four-wave mixing (open circles) versus pump power, with
$b_0 = 10$ and $\Delta = -8 \Gamma$. Lasing threshold (vertical
dashed line) is expected for a reflectivity around $1\%$
(horizontal dashed line), in good agreement with the experimental
data. Inset: Example of a weak-probe reflectivity spectrum.}
\label{fig.FWM}
\end{figure}

We have observed another lasing mechanism when a balanced pumping
scheme using two counterpropagating pump beams F and B is used. In
this configuration FWM appears \cite{Yariv:1977,Abrams:1978}. The
creation of photons in a reflected wave, resulting from a phase
conjugation process, can also be considered as a gain mechanism.
This is reminiscent of optical parametric oscillation where signal
and idler photons are created under a phase matching condition. In
the inset of Fig. \ref{fig.FWM} we show the FWM signal $R_c$
(expressed as the reflection normalized to the incident probe
power) illustrating the narrow spectrum of this phase conjugation
signal. As expected, the maximum gain corresponds to the
degenerate case $\delta=0$ \cite{Lezama:2000}. Thanks to
constructive interference between transmitted and reflected waves,
this mechanism produces huge double-pass gain with cold atoms
\cite{Michaud:2007} and it is thus an efficient mechanism to
trigger laser oscillations \cite{Pinard:1986}. Due to these
interference effects, the threshold for laser oscillation is very
different from the previous cases \cite{Pinard:1986,
Michaud:2007}, and is given by
\begin{equation}
R_c > \left[(1-\sqrt{\tilde{R}})/(1+\sqrt{\tilde{R}})\right]^2= \,
0.9\%
\end{equation}
where $\mathcal{\tilde{R}}=1-\mathcal{L}$. This criterion
(horizontal line in Fig. \ref{fig.FWM}) is well respected for the
threshold of our laser. The output power of this laser is quite
strong (300~$\mu$W), with an energy conversion efficiency of
0.75\% in this case. As two pump beams are used in this situation,
the mechanical effects based on radiation pressure will be
negligible and lasing can be sustained for a long time. However
dipole forces can induce atomic bunching, and change the effective
pump intensity interacting with the atoms \cite{Gattobigio:2006}.

In conclusion, we presented in this Letter three types of laser
using a sample of cold atoms as gain medium. Three different gain
mechanisms were demonstrated as being efficient enough to allow
lasing, even with a low finesse cavity. Comparison between Mollow
and Raman laser shows that the latter has a significantly lower
power, although their gain are of the same order of magnitude.
These two mechanisms can produce high gain at frequencies slightly
detuned from the pump, allowing to distinguish between stimulated
photons from the laser mode and scattered photons from the pump
beam. Thus, they seem to be good candidates for the search of
random lasing in cold atoms, and the combination of these gains
with multiple scattering will be the subject of further
investigations. In addition, the ability to continuously tune from
a Mollow to a Raman laser (by changing the pump detuning), may
allow to study the transformation of transverse patterns from
Laguerre-Gauss to Hermite-Gauss modes \cite{Abramochkin:2004}. The
FWM laser is the most efficient in terms of power, and it should
be possible to study its noise spectrum down to the shot noise
level. This laser has many analogies to an optical parametric
oscillator and seems to be a good candidate to explore non
classical features of light, such as the production of twin beams
\cite{Vallet:1990,McCormick:2007}. Lastly, the coupling between
the cavity mode and the atomic internal and external degrees of
freedom, may also reveal interesting dynamics, especially if a
high-finesse cavity is used
\cite{Courteille:2003,Miller:2005,Ritsch:2006}.

\begin{acknowledgments}
The authors thank G.-L. Gattobigio for his help at the early
stages of the experiment. This work is supported by INTERCAN, DGA
and ANR-06-BLAN-0096.
\end{acknowledgments}


\end{document}